\documentclass[prr, aps, twocolumn, floatfix, superscriptaddress, amsmath,  longbibliography]{revtex4-2}

\usepackage{graphicx}   
\usepackage{dcolumn}    
\usepackage{bm}         

\usepackage{mathtools}
\usepackage{float}
\usepackage{amssymb}

 \usepackage{siunitx}   

\usepackage[dvipsnames]{xcolor}
\definecolor{darkblue}{rgb}{0.0, 0.0, 0.75}
\usepackage{color}
\usepackage[colorlinks=true,
            linkcolor=darkblue,
            urlcolor=darkblue,
            citecolor=darkblue]{hyperref}

	\usepackage[normalem]{ulem} 
	\definecolor{mgreen}{RGB}{1,123,0}

\def \br{{\bf r}}

\def \ms{\text{s}}

\def \ms{\mathrm{s}}

\def \mum{\mu\mathrm{m} }

\usepackage{braket}

\begin{document}

\title{Josephson Dynamics in 2D Ring-shaped  Condensates}

\author{Koon Siang Gan}
\thanks{These authors contributed equally to this work.}
\affiliation{MPA-Q Division, Los Alamos National Laboratories, Los Alamos, New Mexico, 87545, USA}
\author{Vijay Pal Singh}
\thanks{These authors contributed equally to this work.}
\affiliation{Quantum Research Centre, Technology Innovation Institute, Abu Dhabi, UAE}
\author{Luigi Amico}
\affiliation{Quantum Research Centre, Technology Innovation Institute, Abu Dhabi, UAE}
\affiliation{Dipartimento di Fisica e Astronomia "Ettore Majorana", Universit\`a di Catania, Via S. Sofia 64, 95123 Catania, Italy}
\affiliation{INFN-Sezione di Catania, Via S. Sofia 64, 95127 Catania, Italy}
\author{Rainer Dumke}
\affiliation{Centre for Quantum Technologies (CQT), Nanyang Technological University Singapore (NTU), 50 Nanyang Avenue, Singapore 639798, Singapore.}
\affiliation{School of Physical and Mathematical Sciences, Nanyang Technological University, 637371, Singapore}

\date{\today}

\begin{abstract}

We investigate Josephson transport in a fully closed, two‑dimensional superfluid circuit formed by a ring‑shaped $^{87}$Rb Bose–Einstein condensate that contains two optical barriers acting as movable weak links. Translating these barriers at controlled speeds imposes a steady bias current, enabling direct mapping of the current–chemical‑potential ($I$–$\Delta\mu$) characteristics. For narrow junctions ($w \approx 1 \mu\text{m}$) the circuit exhibits a pronounced dc branch that terminates at a critical current $I_{c}=9(1)\times10^{3} \,\text{s}^{-1}$; above this threshold the system switches to an ac, resistive regime. Classical‑field simulations that include the moving barriers quantitatively reproduce both the nonlinear $I$–$\Delta\mu$ curve and the measured $I_{c}$, validating the underlying microscopic picture. Analysis of the ensuing phase dynamics shows that dissipation is mediated by the nucleation and traversal of vortex–antivortex pairs through the junctions, while the bulk condensate remains globally phase‑locked—direct evidence of the ring’s topological constraint enforcing quantized circulation. These results establish a cold‑atom analogue of a SQUID in which Josephson dynamics can be resolved at the single‑vortex level, providing a versatile platform for atomtronic circuit elements, non‑reciprocal Josephson devices, and on‑chip Sagnac interferometers for multi‑axis rotation sensing. 
\end{abstract}

\maketitle



\section{Introduction}\label{sec:intro}
Josephson dynamics occurs when two phase coherent systems are separated by a thin barrier \cite{josephson1962possible}. Originally discovered in superconducting systems, Josephson transport  is characterized by a non-linear current-phase relation and a critical current $I_c$. 
For currents below $I_c$, dissipationless Cooper pair flow occurs, defining the dc Josephson effect.
When the current exceeds $I_c$, a voltage can develop across the barrier due to the tunneling of quasi-particles (broken Cooper pairs), leading to the ac Josephson effect. This foundational phenomenology underpins both a wide range of cutting-edge technologies and ongoing fundamental research \cite{barone1983josephson,tinkham2004introduction}.

In contrast to point-like junctions, extended two-dimensional (2D) long junctions, where the phase distribution across the superconducting leads cannot be neglected, exhibit qualitatively different Josephson dynamics. 
In these systems, the spatial structure of the phase plays a critical role, modifying key features 
such as the critical current, the current-phase relation, and quasiparticles behavior.
When subjected to an external magnetic field, such junctions exhibit rich phenomena 
including the formation of Josephson vortices, 
nonreciprocal (diode-like) transport, and Fraunhofer interference patterns \cite{barone1983josephson,rashidi2025self,fedorov2012nonreciprocal,ozyuzer2007emission}. 

Spatially closed coherent circuits add an important  dimension to Josephson dynamics: 
they enable macroscopic quantum superpositions of clockwise and anticlockwise matterwave  currents  \cite{leggett1987macroscopic,leggett1980macroscopic}. On the  technological applications side, such macroscopic superpositions, turning out to be  exquisitely sensitive to magnetic flux change,  are the central feature exploited in superconducting quantum interference devices (SQUIDs) \cite{khalid2010}.

\begin{figure*}
\includegraphics[width=1.0\linewidth]{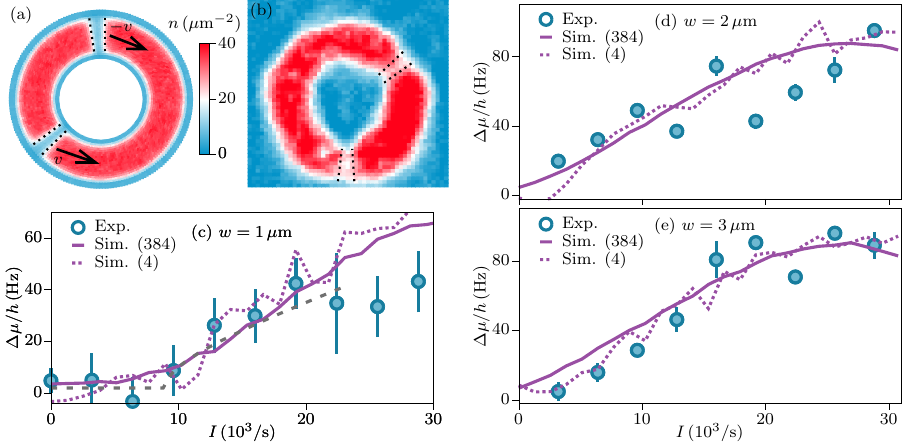}
\caption{Superfluid-resistive regimes. 
(a) Simulation of a homogeneous 2D ring condensate consisting of two mobile tunnel barriers of height $V_0$ and width $w$. 
The barriers move at a constant velocity $v$ (arrows) from their initial positions in the opposite direction to induce a net current $I$. 
(b) In situ absorption image taken after barrier translation for $v/v_0 = 0.4$ ($v_0 =  2\pi \hbar/(m R)$) with a barrier width of $3\mum$. $z$ is determined by subtracting the atom number of the larger sector by the atom number of the smaller sector at the end of the movement. $z_0$ is determined similarly by loading the BEC into the potential where the barrier is at its end position. Shot-to-shot total atom number and relative atom number fluctuations in the 2 arms of the ring are within $10\%$.
(c) Measurements of the current-chemical potential relation (dots) for $w=1\, \mum$, alongside the simulation results (continuous and dotted lines) that are obtained with barrier height $V_0/\mu=1.75$ at temperature $T=62\, \mathrm{nK}$,  
where $\mu$ is the mean-field energy. 
Simulations averaged over 4 realizations exhibit visible statistical fluctuations, which are suppressed when increasing the sampling to 384 realizations. 
We fit the results with $\Delta \mu = \sqrt{I^2 -I_m^2}/G$ (dashed line) to determine the threshold current $I_m$ and the conductance $G$.
(d, e) Current-chemical potential relations for barrier widths $w=2\, \mum$ and $3\, \mum$, the barrier heights are both $V_0/\mu=1.4$. Each experimental data point in panels (c-e) is obtained by averaging over 2 to 3 realizations.
}
\label{Fig:system}
\end{figure*}


Ultracold atomic gases have emerged as powerful quantum simulators of mesoscopic circuits, spawning the field of \emph{atomtronics} \cite{Amico2022, Polo2024}. 
In particular, Bose--Einstein condensates (BECs) separated by a barrier provide
phase‑coherent neutrally charged matter waves analog to superconducting 
Josephson junctions, but with an enhanced flexibility and 
control of the system's parameters \cite{Smerzi1997,Kwon2020,Luick2020}.  
We emphasize that the current experimental know-how of the cold atoms field allows to monitor the actual {\it microscopic dynamics} of the system. Therefore, in addition to their own interest, atomtronic circuits have been providing analog simulators of driven superconducting circuits in which the aforementioned dynamics happen at so fast time scales that it is very difficult if not just impossible to be accessed \cite{SinghShapiro, DelPace2025, Bernhart2025, Singh2025}.  

In atomic systems,  Josephson dynamics is achieved by creating a finite relative velocity between barrier and condensate,  either to move the condensate with a fixed barrier or by moving the latter with the condensate at rest. Josephson oscillations and self trapping \cite{Smerzi1997} were first observed in a double‑well BEC that realised a single bosonic junction
\cite{albiez2005}, followed by the demonstration of \emph{a.c.} and \emph{d.c.} Josephson effects and current-phase relations in a single weak link \cite{Levy2007,Kwon2020, Luick2020, Pace2021, Singh2025Weak}.  
Ring‑shaped condensates with movable barriers have since enabled closed‑loop superfluid circuits that exhibit quantized persistent currents \cite{PhysRevA.86.013629, Campbell2011, Pace2022, Cai2022}, critical-current behavior and vortex‑mediated phase slips \cite{Wright2013, Campbell2014, Piazza2009, Eckel2014, Mathey2014, Mathey2016}, 
and flux-dependent interference patterns 
\cite{Ryu2013, Ryu2020, Kiehn2022} (see also \cite{Polo2025} for a recent review,  and \cite{Gorg2025} for implementation with a rotating-box potential). 

Previous ring‑BEC experiments have established key building blocks of atomtronics, including quantized persistent currents, critical currents of tunnel junctions in toroidal traps, phase slips and hysteresis in rotating weak links, and resistive flow in source–drain ring geometries. In a notable pioneering work, $I-\Delta \mu$ characteristics have been measured in a 3D toroidal BEC \cite{Campbell2014}. Here we advance these studies by directly measuring the transport characteristic $I-\Delta \mu$ of a fully closed, two‑junction, quasi‑2D ring circuit, and by using junction‑width control to access a pronounced dc branch with a sharp threshold (narrow junctions) versus an immediately resistive response (wider junctions). Parameter‑matched classical‑field simulations reproduce the measured nonlinear $I-\Delta \mu$  curve and link the resistive branch to vortex–antivortex phase slips localized at the junctions while the bulk remains globally phase‑locked, highlighting the distinct role of the ring’s topological constraint in Josephson dynamics.


Here we study the Josephson dynamics in 2d quantum‑degenerate superfluids.
 Our system is trapped   by a 
digital micromirror device (DMD) \cite{gauthier2016direct,gauthier2021chapter}.  Two
repulsive ``optical paddles'' positioned diametrically across the ring are translated synchronously, forming a pair of moving weak links that act as Josephson junctions.  By varying the paddle speed we map the current–phase relation, identify a well‑defined critical current, and observe a transition from phase‑locked, lossless transport to dissipative dynamics accompanied by quantised vortex production.  Because the intrinsic time scales of the atomic circuit lie in the millisecond regime—orders of magnitude slower than in  superconductors—phase evolution, phase slips,
and vortex nucleation can be imaged in real time.  Our results
provide a microscopic view of Josephson dynamics that is
inaccessible in conventional electronic devices and underline the aforementioned potential of atomtronic platforms for emulating complex quantum transport phenomena.


\section{Experiment setup} 

A pure $^{87}$Rb BEC in the $\ket{F=2,m_F=+2}$ state is created in a 1064nm crossed dipole trap with $10^5$ atoms. This is transferred into a combined 532nm blue-detuned trap, consisting of a $7.5\mu m$ sheet trap for axial confinement and a configurable radial trap generated by a DMD (see Appendix \ref{sec:App:details} for details). 

An initial $25\mu m$ diameter disc was chosen for its shape convenience while maximising the loading efficiency up to 80$\%$. By updating the images displayed on the DMD at some frame rate, the optical potential can be adiabatically changed. This disc evolves into a $25\mu m$, $6\mu m$ thick ring within 36ms at $0.7\mu $m/ms and the condensate in this combined trap is held constant for 500ms for the system to relax with $10^4$ atoms \cite{Aidelsburger2017,Pezze2024}. The transverse trap parameters are $\omega_z = 2\pi \times 1.2$ kHz, its oscillator length $l_z = \sqrt{\hbar/(m\omega_z)} = 0.31\mu$m, and the speed of sound $ c_0 = \sqrt{n_sg_{2D}/m} \approx 1.1mm/s$ for $n_s = 28 (\mum)^{-2}$,  $g_{2D} = \sqrt{8\pi}\hbar^2a_s/(ml_z) $ is the 2D interaction parameter \cite{Aidelsburger2017,Sunami2022}, $a_s = 5.3$nm is the scattering length, $\hbar$ is the reduced Planck constant, and $m$ is the mass. The speed of sound sets the evolution speed limit of the optical potentials without introducing excitations. Radial ring frequencies of such thickness are typically tens of Hz \cite{Campbell2011,Aidelsburger2017,Cai2022}, thus $\mu_{3D}/h \approx 600$ Hz $< \omega_z/(2\pi) = 1.2$ kHz, putting the system in the quasi-2D regime \cite{Morizot2006}. The healing length of our system is $\xi_{2D} = \frac{1}{\sqrt{2\tilde g n_s}} = 0.46\mu$m , where $\tilde g = \sqrt{8\pi} \frac{a_s}{l_z}$  \cite{Zoran2011} .

 Investigation of the dynamics is initiated with the introduction of two repulsive optical barriers within 20ms. The barriers then move azimuthally towards each other through a series of evolving frames. Different angular speeds can thus be achieved by adjusting the frame duration and the number of frames to ensure the speed of sound is not exceeded. The starting paddle positions are constant for all angular speeds $v$ to maintain a consistent initial condition, hence the final positions of the barriers differ for varying $v$. Finally, the atoms are imaged in situ using absorption imaging and the optical depth computed. The imbalance $\Delta z = (N_1 - N_2)/(N_1 + N_2)$ is determined through the atom number difference between the 2 ring sectors. Subtracting the imbalance without the barriers' rotation, $z_0$, provides meaningful information on the dynamics.

\begin{figure*}[t]
\includegraphics[width=1.0\linewidth]{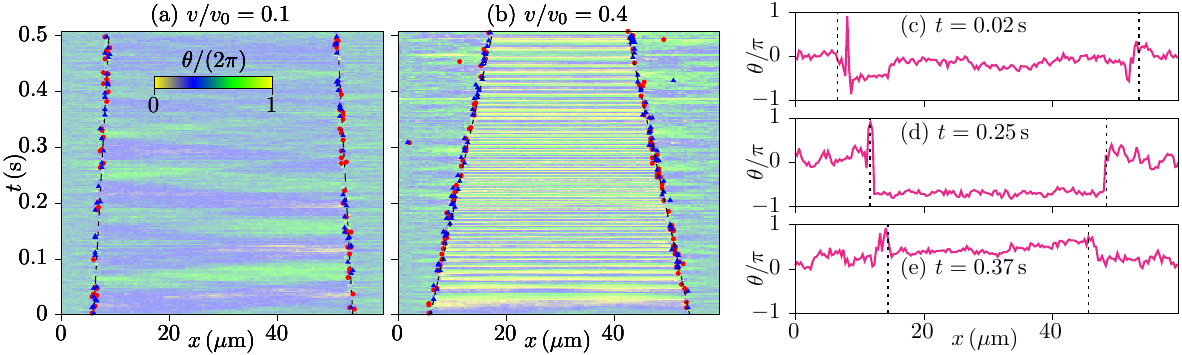}
\caption{Dissipation mechanism. Time evolution of the phase $\theta(x, L_y/2)$ in the $x$ direction of a single sample for barrier velocities (a) $v/v_0=0.1$ and (b) $v/v_0=0.4$, where $v_0$ is the characteristic velocity. Vortices (triangles) and antivortices (dots) are calculated using the phase change near the line at $L_y/2$ and the dotted lines denote the positions of moving barriers. 
(c-e) $\theta(x, L_y/2)$ at different times $t$, from the time evolution shown in panel (b), 
where vertical dashed lines mark the barrier positions. 
}
\label{Fig:phase}
\end{figure*}
%

\section{Simulation method}
We simulate the system dynamics using a classical-field method within the truncated Wigner approximation, which is a well-established framework for modelling condensate dynamics in ultracold atom experiments \cite{Blakie2008, Polkovnikov2010, Singh2017, Singh2021}. 
Within the classical-field approximation, we replace the operators $\hat{\psi}$ in the Hamiltonian and equations of motion by complex numbers $\psi$. The initial states $\psi (\br, t=0)$ are sampled in a grand canonical ensemble with chemical potential $\mu$ and temperature $T$ via a classical Metropolis
algorithm. We then propagate each state using the equations of motion
\begin{align}\label{eq:eom}
 i \hbar \dot{\psi}(\br, t) = \Bigl(  - \frac{\hbar^2}{2m} \nabla^2 + V(\br, t) + g|\psi|^2 \Bigr) \psi(\br, t),
\end{align}
where $g$ is the repulsive interaction. The barrier potential is given by 
$V({\bf r},t)  = V_0 (t)  \sum_i \exp \bigl[- 2 \bigl( x-x_i(t) \bigr) ^2/w^2 \bigr]$, 
where $V_0(t)$, $w$ and $x_i(t)$ are the strength, width and location.
$i=1, 2$ denote the two barriers with $x_i(t) =v_i t + x_{0, i}$, 
where $v_i$ is the velocity and $x_{0, i}$ is the initial location. 
Similarly to the experiments, the two barriers move at a constant velocity $v$ in the opposite direction for a fixed duration of $0.5\, \ms$, see Fig. \ref{Fig:system}(a). 
For numerical simulations, we map the ring geometry onto a rectangular strip consisting of $200 \times 23$ lattice sites, with a discretization length of $l=0.3\, \mum$. 
We gradually ramp up $V_0$ over $0.1\, \ms$. After moving both barriers, we determine the imbalance $\Delta z$ from atom number difference between the reservoirs separated by two barriers, allowing us to determine $\Delta z$ as a function of $v$. The net induced current is given by $I= v N/L_x$, where $N$ is the total atom number and $L_x$ is the strip length. $\Delta z$ is connected to the chemical-potential difference $\Delta \mu = N E_c \Delta z/2$, where $E_c = 4 (\partial \mu/\partial N)$ is the charging energy. For homogeneous clouds, this results in $\Delta \mu = 2 \mu \Delta z$, where $\mu= g n$ is the mean-field energy.

\section{DC-AC Josephson effects}
In Fig. \ref{Fig:system}(c), we present measurements of the $I-\Delta \mu$ relation for a barrier width of $w= 1 \, \mum$. The observed behavior is characteristic of atomic Josephson junctions \cite{Levy2007, Kwon2020, Pace2021}. Notably, above a certain threshold current, the measured $\Delta \mu$ increases nonlinearly, indicating a transition from superfluid to resistive transport. This nonlinearity serves as a hallmark of the critical current separating dissipationless and dissipative regimes. We fit the response with the prediction of the RSJ (resistively shunted junction) circuit model, i.e., $\Delta \mu = \sqrt{I^2 -I_m^2}/G$, with the conductance $G$ and the threshold current $I_m$ as the fitting parameters \cite{SinghShapiro}.
This gives $I_m= 9 \pm 1.2\, (10^3/\ms)$ and $\hbar G=86$, where $I_m$ is twice the critical current of individual junction. 
To benchmark the experiments, we perform simulations using the same system parameters and show these results for the barrier height $V_0/\mu=1.7$ and width $w=1\, \mum$ in Fig. \ref{Fig:system}(c), which describes the nonlinear behavior well, see also Appendix \ref{sec:App:comp}. From the simulations, we obtain $I_{m, \ms}= 9.7 \pm 0.4\, (10^3/\ms)$ and $\hbar G_\ms=75$, which are in  agreement with the experimental values. 

As the junction tunneling is controlled by its parameters, we can explore strong resistive regimes by increasing the barrier width, which suppresses the supercurrent tunneling \cite{Singh2020jj}. 
In Figs. \ref{Fig:system}(d, e), we present measurements for barrier widths $w= 2$ and $3\, \mum$. 
For both widths, there is no dissipationless regime as the junctions immediately enter the resistive regime at any nonzero value of the bias current. This observation is consistent with the simulation results shown for the corresponding widths at $V_0/\mu=1.4$ in Figs. \ref{Fig:system}(d, e).

\section{Dissipation dynamics}
To gain physical insights, we examine the underlying phase dynamics in our simulations. 
We calculate the phase distribution $\theta(x, L_y/2)$ of a single sample of the system,  
where $\theta(x, y)$ is the phase field of $\psi(x, y)$.  
Figs. \ref{Fig:phase}(a, b) show the time evolution of $\theta(x, L_y/2)$ in both the dissipationless ($v/v_0=0.1$) and resistive ($v/v_0=0.4$) regimes. 
The characteristic velocity is defined as $v_0 =  2\pi \hbar/(m R) \approx 0.37\, \mathrm{mm/s}$, where $R$ is the ring radius. 
Due to the high potential barrier (relative to the mean-field energy), low density regions develop at the barrier location. These regions lead to enhanced phase fluctuations and the nucleation of vortex excitations.
We detect vortex excitations by computing the phase winding around a lattice plaquette of size $l\times l$ using $\sum_{\Box} \delta \theta(x,y) = \delta_x \theta(x,y) + \delta_y\theta(x+l,y)+\delta_x\theta(x+l,y+l)+\delta_y\theta(x,y+l)$, 
where the phase differences between sites are taken to be $\delta_{x/y} \theta(x,y)  \in (-\pi, \pi]$. 
A net phase winding of $2\pi$ ($-2\pi$) is identified as a vortex (antivortex).
For $v/v_0=0.1$, phase fluctuations in the bulk remain weak, as expected in a superfluid regime. 
In contrast, at $v/v_0=0.4$, corresponding to the resistive regime, 
strong phase fluctuations are observed near the barrier. 
This is illustrated in Figs. \ref{Fig:phase}(c-e), which show clear instances of phase slips localized at the junction.  
These phase slips lead to the nucleation of vortex-antivortex pairs that propagate in the bulk during the evolution, see Appendix \ref{sec:App:dissipation} for details. 
Interestingly, despite the localized phase drops at the barriers, the phase in the bulk region between the barriers remains coherent, indicating a dynamically phase-locked state, a hallmark  of the coherent Josephson effect in ring geometries.

\begin{figure}[t]
\includegraphics[width=1.0\linewidth]{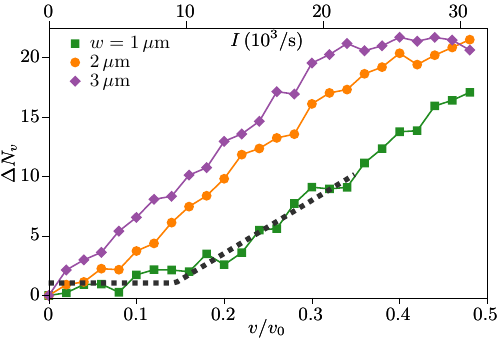}
\caption{Relative vortex number $\Delta N_v = N_v - N_v(v=0)$ as a function of velocity $v/v_0$ and  current $I$ (second axis) for barrier widths $w=1$, $2$ and $3\, \mum$, corresponding to the simulation results presented in Fig. \ref{Fig:system}. 
The fit to the bilinear function (dashed lines) yields the threshold velocity $v_m/v_0=0.14 \pm 0.02$ or threshold current $I_m= 8.9 \pm 1.3\, (10^3/\ms)$.
}
\label{Fig:vortex}
\end{figure}

By counting all vortices and antivortices across the entire sample, we determine the total vortex number $N_v$ and average it over the initial ensemble.
Fig. \ref{Fig:vortex} presents the relative vortex number at varying $v/v_0$, 
using the same system and barrier parameters as in Fig. \ref{Fig:system}.
For larger barrier widths ($w=2$ and $3\, \mum$), $N_v$ increases with velocity, showing a gradual saturation at higher $v$. This trend closely follows the behavior of the imbalance observed in Figs. \ref{Fig:system}(d,e). A similar pattern is observed for the narrower barrier width  ($w=1\, \mum$), where the vortex number increases in a manner consistent with the imbalance growth. Notably, the vortex number exhibits two distinct regimes: at low velocities, 
$N_v$ remains nearly constant, while at intermediate and high velocities, it rises sharply.  We fit the vortex response with a bilinear function to determine the threshold velocity for significant vortex generation, finding $v_m/v_0=0.14 \pm 0.02$.

\section{Summary and outlook}
We directly measured the nonlinear $I$--$\Delta\mu$ response of a fully closed, two-junction, two-dimensional ring Bose--Einstein condensate by translating optical weak links at controlled speeds. For narrow junctions ($w=1~\mu\mathrm{m}$), we observed a pronounced dc branch that terminates at a threshold current $I_m = (9.0 \pm 1.2)\times 10^{3}~\mathrm{s^{-1}}$, whereas wider junctions ($w=2,3~\mu\mathrm{m}$) enter the resistive regime immediately for any nonzero bias. An RSJ-type fit yields conductances consistent with classical-field simulations that include the moving barriers; the simulations quantitatively reproduce both the measured nonlinearity and the threshold $\big(I_{m,s}=(9.7 \pm 0.4)\times 10^{3}~\mathrm{s^{-1}};\ \hbar G=86~\text{(experiment)},~\hbar G_s=75~\text{(simulation)}\big)$. We demonstrate that the dc-ac Josephson regimes in the system correspond to specific {\it microscopic dynamics}: the  dissipation is found to be mediated by the nucleation and traversal of vortex--antivortex pairs localized at the junctions, while the bulk between junctions remains phase-coherent. 
Counting vortices across the sample yields a threshold $v_m/v_0=0.14(2)$ with $v_0=2\pi\hbar/(mR)$, matching the observed transport onset. 
Despite the localized phase drops at the junctions in the resistive regime, 
the system self-adjusts to preserve global phase continuity—a hallmark of both atomic and superconducting Josephson circuits in closed-loop configurations \cite{Wright2013,Campbell2014,Ryu2013, Campbell2011,Ryu2007,Ryu2020, Kiehn2022, Amico2022}. 

Our study confirms ultracold atoms as a powerful microscopic testbed for studying closed‑loop Josephson dynamics and exploring new regimes of coherent matter-wave transport.
In particular, three immediate directions follow. 
\textit{Josephson‑vortex and diode physics:} Asymmetric double barriers or synthetic gauge fields could create non‑reciprocal superflow, realizing a matter‑wave analogue of the superconducting Josephson diode and adding a unidirectional “rectifier” to the atomtronic toolbox \cite{Ustinov1998,Pal2022,Wu2022,fedorov2012nonreciprocal,Gou2020,Lau2023}. \textit{Precision rotation sensing:} The ring is a natural Sagnac interferometer; its dual junctions provide built‑in splitting and recombination, suggesting a compact gyroscope sensitive to sub‑Earth‑rate rotations. Arrays of orthogonal planar rings could yield full three‑axis inertial read‑out for navigation and geodesy \cite{Gustavson1997,Dickerson2013,Canuel2006,Ryu2020,Krzyzanowska2022}. \textit{Complex atomtronic circuitry:} Coupling several rings or embedding them in lattices would emulate flux‑biased SQUID networks, phase batteries, and topological qubits, all observable in situ \cite{Strambini2020,Mooij1999,Ioffe2002,Amico2022}.

With continued advances in laser sculpting, low‑noise detection, and long coherence times, ultracold‑atom Josephson circuits will progress from proof‑of‑principle demonstrations to versatile quantum simulators and precision sensors, bridging condensed‑matter electronics and neutral‑atom technology \cite{gauthier2016direct,Ryu2020,Kiehn2022,Amico2022}.

\section*{Acknowledgments}
We gratefully acknowledge fruitful discussions with T. Haug, L. C. Kwek, and J. Polo.
We acknowledge funding from NRF via NRF2020-NRF-ISF004-3515 and  NRF2021-QEP2-03-P01.
This project is supported by the National Research Foundation, Singapore through the National Quantum Office, hosted in A*STAR, under its Centre for Quantum Technologies Funding Initiative (S24Q2d0009).


\bibliography{references}


\appendix
\renewcommand{\thefigure}{A\arabic{figure}}
\renewcommand{\theequation}{A\arabic{equation}}
\setcounter{figure}{0}
\setcounter{equation}{0}

\section{Experiment details} \label{sec:App:details}

\begin{figure}
\includegraphics[width=1\linewidth]{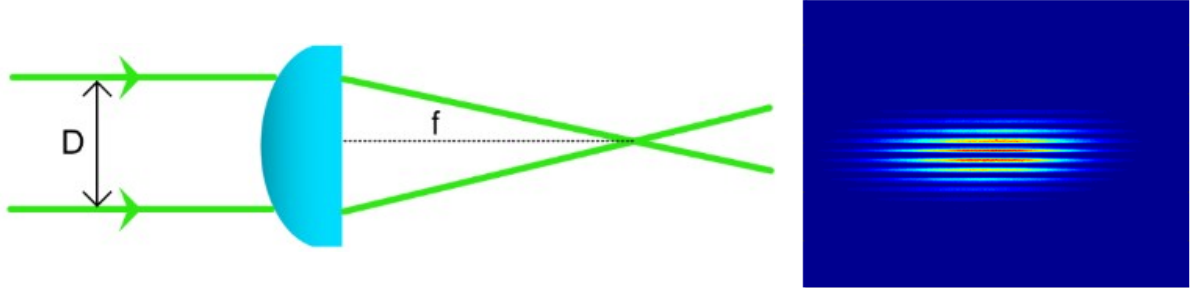}
\includegraphics[width=0.5\linewidth]{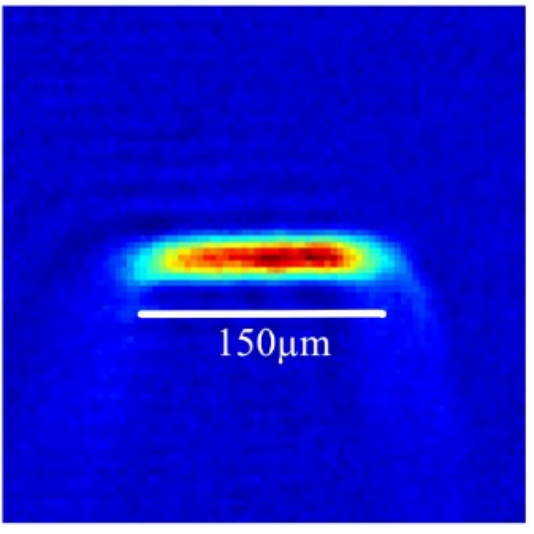}
\caption{(Top) The interference of 2 beams generates sheets of lattice with lattice spacing $d \approx \frac{f\lambda}{D}$, with a background profile of a  beam intensity at the focus of a lens. The false colour image on the right shows the optical profile of our sheet trap at the focus. Although there are multiple sheet sites available, only 2 sheets are suitable to hold against gravity if the correct laser power is chosen. (Bottom) An image of a BEC loaded into one of the sheet sites and held for some time. The length of the sheet is determined by the beam parameters at the focus and the laser power. The wings at the edge of the condensate are atoms falling out of the sheet.
}
\label{AppFig:sheet}
\end{figure}

\begin{figure}
\includegraphics[width=0.75\linewidth]{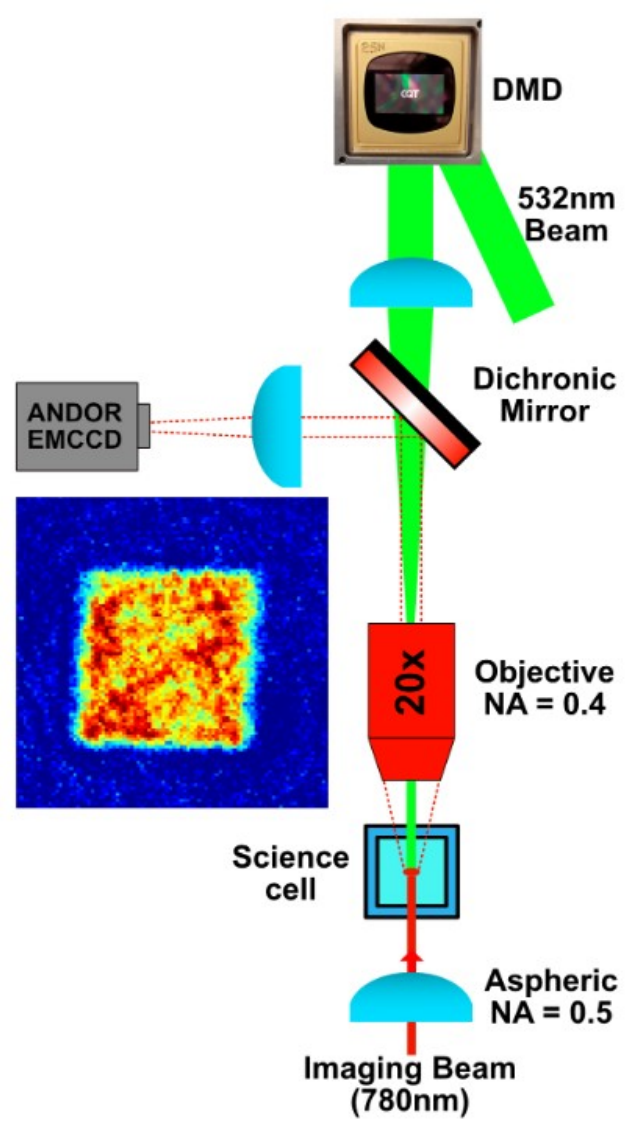}
\caption{Experiment diagram of the DMD trap and the imaging system. A dichroic mirror enables the imaging of the BEC using a single objective lens for high resolution imaging. An imaging beam is delivered from the bottom of the science cell to perform absorption imaging. The image on the side is an in-situ image of a BEC loaded into a $30\mu m$ x $30\mu m$ square.
}
\label{AppFig:dmd}
\end{figure}

The combined 532nm blue-detuned trap consists of a sheet trap for axial confinement and a configurable radial trap generated by a DMD. The sheet trap is created through interference of 2 collimated beams focused to the atoms with a single lens (Fig. \ref{AppFig:sheet}). The beam waists at the focus of the 300mm lens are $\omega_x = 71\mu m , \omega_y = 30\mu m$, and the interference sheets spacing are $d = f\lambda/D = 7.5\mu$m, where $f$ is the focal length and $D$ is the distance between the 2 collimated beams. The choice of beam waists and beam power (2 x 60mW) ensures only 2 possible sheet sites can be loaded into. The DMD (Vialux DLP9500) optical setup (Fig. \ref{AppFig:dmd}) delivers dynamic optical potentials and is digitally controlled with a self made user interface. A 532nm 10mm diameter collimated beam ($1/e^2$) is reflected off the surface of the DMD and demagnified initially by 4x and then 20x with a 200mm lens and a NA $ = 0.40$ objective (Mitutoyo M Plan APO NIR 20x), which projects the DMD image onto the BEC. Accounting for losses, the diffraction efficiency of the 1st order of the DMD, and the chosen trap geometry, up to 50mW of power can be delivered to the atoms. Imaging of the BEC is performed through absorption using the same 200mm - objective lens but with a dichroic mirror that reflects 780nm light towards a 1x telescope and captured by a Princeton Instruments CMOS camera. The resolution of our imaging system is 1$\mu$m, while the DMD trap has a resolution better than 1$\mu$m because it has a shorter wavelength.

A 3-stage optical evaporation sequence creates a pure BEC of approximately 1 x $10^5$ atoms in 7s in an optical Crossed Dipole Trap 
(CDT) generated by a 1064nm laser. Transfer of the BEC starts by simultaneously ramping up the combined blue trap at their respective maximum laser powers within 100ms while keeping the 1064nm CDT on. The CDT is then ramped down within 10ms and the combined trap is held constant for 500ms to allow for the condensate to settle. The transfer efficiency can be maximised to almost 80$\%$ by choosing a trap pattern that matches the size of the BEC in the CDT and a 25$\mu$m diameter disk fulfills this condition. The disk then evolves into a $25\mu m$, $6\mu m$ thick ring in 36ms at 0.7$\mu$m/ms, below the speed of sound, through a loaded set sequence of images. Finally, the ring BEC is held constant for 500 ms to allow the condensate to settle. A separate 25ms TOF was performed after each transfer sequence and each trap potential evolution to check for heating of the BEC. A condensate without any thermal component and an inversion of its aspect ratio confirm that no heating has occurred. These TOF images were acquired by absorption imaging from a plane perpendicular to gravity. 

All barrier rotation experiments have the same starting position for both barriers, implying different end positions for different rotation speeds. Both barriers are introduced within 20ms at 1ms/frame and they move towards each other azimuthally at the desired velocities. In every image both sectors have to be defined by $N_1$ and $N_2$ for consistency, an image mask is defined to isolate each sector and its pixel value sum computed. The imbalance $\Delta z = (N_1-N_2)/(N_1+N_2)$ is calculated by the difference of the sum values of the sectors. Imbalance without rotation $z_0$ is determined similarly by loading the BEC at the barriers' end positions for each velocity.

\section{Experiment and simulation comparison} \label{sec:App:comp}

The simulation was performed by assuming a constant potential barrier height with varying widths. In our experiment, a constant intensity profile is created from the incident Gaussian through a half-toning technique. This technique enables grayscaling for a binary device like the DMD. A 4x4 Bayer matrix is applied to the incoming Gaussian beam to flatten the intensity profile as much as possible. This flattened intensity profile is convolved with the desired image to generate the desired optical potential. Experimentally, this is done by multiplying the desired image like a ring, by a half-toned image of a Gaussian beam. The end result is that the BEC sits in the region with a constant potential height throughout the region of interest rather than variable heights at different areas of the trap. 

For barrier widths (2$\mu$m and 3$\mu$m) larger than the resolution limit of our DMD optical potential system (1$\mu$m), the potential height of the barrier remains constant but the barrier width varies, this is as per the simulations. However for barrier widths close to the resolution limit, both the potential height and barrier width are coupled to each other. This creates discrepancies to the numerical model at higher rotation speeds as seen in Fig. \ref{Fig:system}(c).

\begin{figure}
\includegraphics[width=1.0\linewidth]{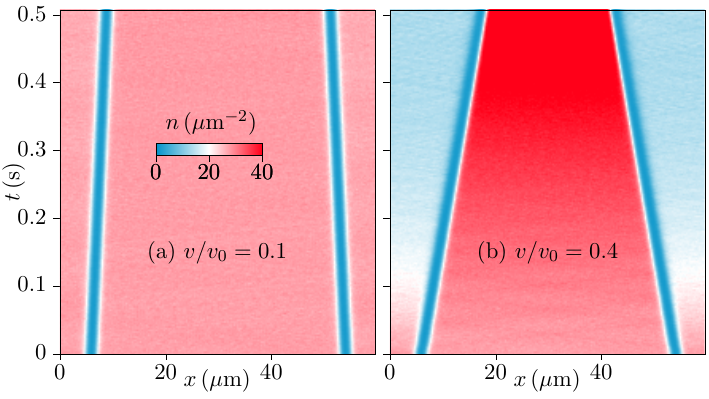}
\caption{(a, b) Time evolution of the averaged density profile $n(x)$ for barrier velocities 
(a) $v/v_0=0.1$ and (b) $v/v_0=0.4$. 
}
\label{AppFig:den}
\end{figure}

\begin{figure}
\includegraphics[width=1.0\linewidth]{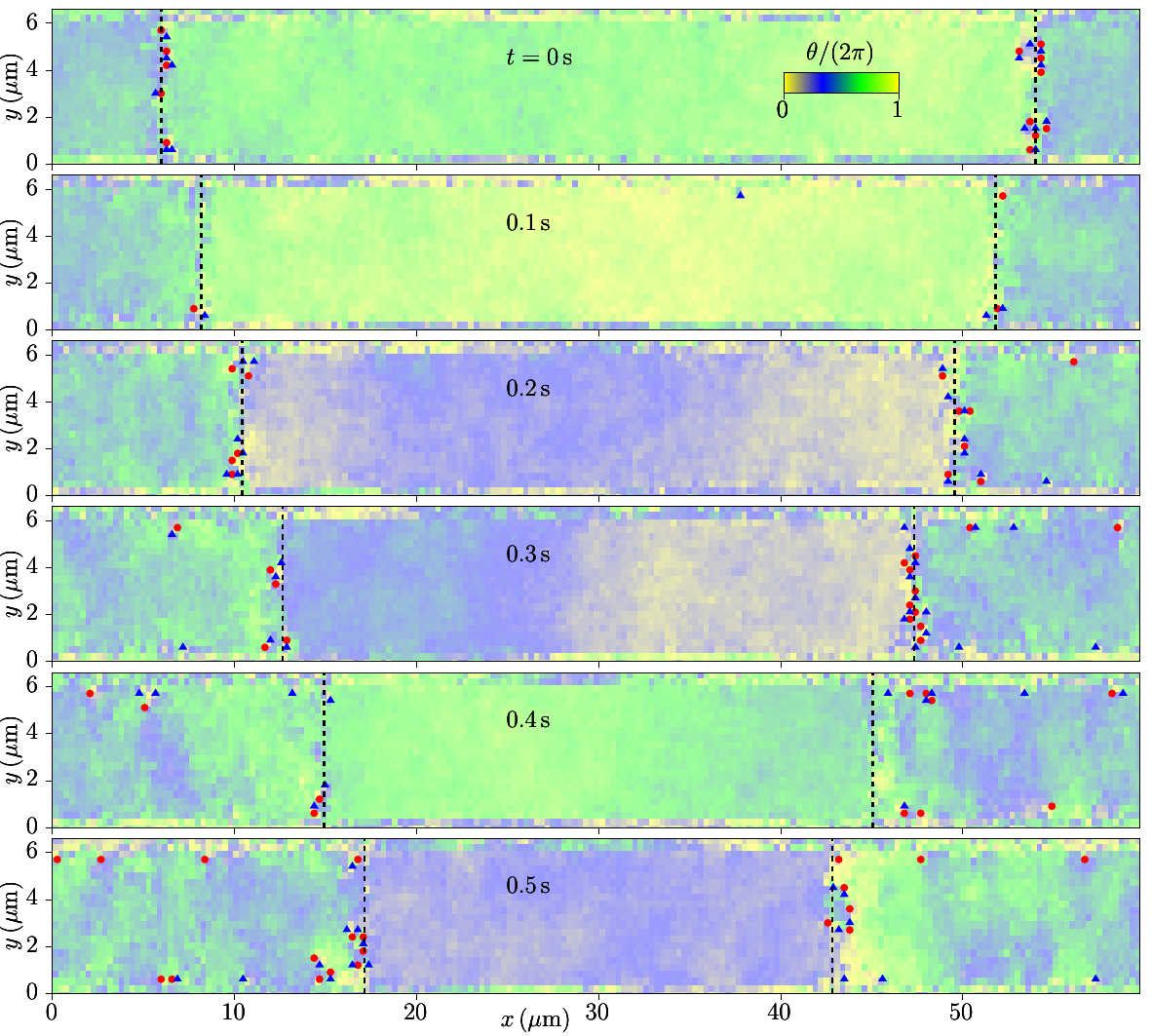}
\caption{Snapshots of the phase profile $\theta(x,y )$ of a single trajectory for the barrier velocity $v/v_0=0.4$, at times $t=0$, $0.1$, $0.2$, $0.3$, $0.4$, and $0.5\, \mathrm{s}$.
During the evolution, the barrier motion (vertical dashed lines) results in the nucleation of vortex pairs that propagate in the bulk.
Vortices and antivortices are depicted as red dots and blue triangles, respectively. 
}
\label{AppFig:nv2d}
\end{figure}

\section{Dissipation dynamics} \label{sec:App:dissipation}
Here, we elaborate on the dissipation dynamics of the density and phase in the simulation. 
Figure \ref{AppFig:den} shows the evolution of the averaged density $n(x)$ along the $x$ direction for the same parameters as in Fig. \ref{Fig:phase}. 
In the superfluid regime ($v/v_0=0.1$), the barrier motion does not cause any noticeable change in the density across the barrier, consistent with dissipationless flow. 
When the barrier velocity exceeds the critical velocity, however, the barrier motion induces a  build-up of density in the bulk between the barriers. 
This density accumulation is accompanied by phase slips, which act as the microscopic mechanism of dissipation: vortex-antivortex pairs are nucleated at the barriers and shed into the condensate, thereby converting coherent superflow into excitations, see Fig. \ref{AppFig:nv2d}.
As a result, the system relaxes through energy redistribution, signaling the onset of dissipative transport.

\end{document}